\documentclass[%
 reprint,
 amsmath,amssymb,
 prl,
 showpacs,
]{revtex4-1}

\usepackage[capitalise]{cleveref}
\usepackage{graphicx}% Include figure files
\usepackage{dcolumn}% Align table columns on decimal point
\usepackage{bm}% bold math
\usepackage{amsmath}
\usepackage{upgreek}
\usepackage{color}
\usepackage{natbib}
\newcommand{\micron}{\mu\mathrm{m}}
\newcommand{\second}{\mathrm{s}}

\newcommand{\ghom}{g_0}

\begin{document}

\preprint{APS/123-QED}

\title{Critical Point in Self-Organized Tissue Growth}

\author{Daniel Aguilar-Hidalgo$^{1,2}$}
\author{Steffen Werner$^{1,3}$}\altaffiliation[Present address: ]{AMOLF, Science Park 104, 1098 XG Amsterdam, The Netherlands.}%
\author{Ortrud Wartlick$^2$}%
\author{Marcos Gonz\'alez-Gait\'an$^2$}%
\author{Benjamin M. Friedrich$^{1,3}$}%
\author{Frank J\"ulicher$^{1,4}$}\email{julicher@pks.mpg.de}%
\affiliation{$^1$Max Planck Institute for the Physics of Complex Systems, N\"othnitzer Stra\ss e 38, 01187 Dresden, Germany}%
\affiliation{$^2$Department of Biochemistry, Faculty of Sciences, University of Geneva, 1205 Geneva, Switzerland}
\affiliation{$^3$cfaed, TU Dresden, 01062 Dresden, Germany}
\affiliation{$^4$Center for Systems Biology Dresden, Pfotenhauerstraße 108, 01307 Dresden, Germany}

\begin{abstract}
We present a theory of pattern formation in growing domains inspired by biological examples of tissue development.
Gradients of signaling molecules regulate growth, while growth changes these graded chemical patterns by dilution and advection.
We identify a critical point of this feedback dynamics, which is characterized by spatially homogeneous growth and proportional scaling of patterns with tissue length. 
We apply this theory to the biological model system of the developing wing of the fruit fly \textit{Drosophila melanogaster} and quantitatively identify signatures of the critical point.

\end{abstract}



\maketitle

How tissues grow to their correct size and become spatially patterned during development is a key question in biology. 
Specific signaling molecules, called morphogens, control tissue patterning and growth \cite{Martin:2004jx,Wartlick:2011by,Restrepo:2014fw}. 
These morphogens are locally produced and secreted. They spread in the target tissues, 
where they form long-range graded concentration profiles \cite{Turing:1952vn,Wolpert:1969wu,Meinhardt:1994,SimpsonBrose:1994ud,Lawrence:1996tm,Jaeger:2004ko,Julicher:2005va,Wartlick:2009jg,Kicheva:2007bha,Muller:2013it,AguilarHidalgo:2013bw}.
Control of tissue growth by morphogens implies a self-organized feedback between growth and chemical gradients, whereby morphogen profiles instruct tissue growth, while growth in turn feeds back on these chemical gradients, e.g. by advection and dilution of morphogens. This mutual coupling between the dynamics of morphogen profiles and tissue growth is still poorly understood.

\begin{figure}[t!]
\includegraphics[width=0.92\linewidth]{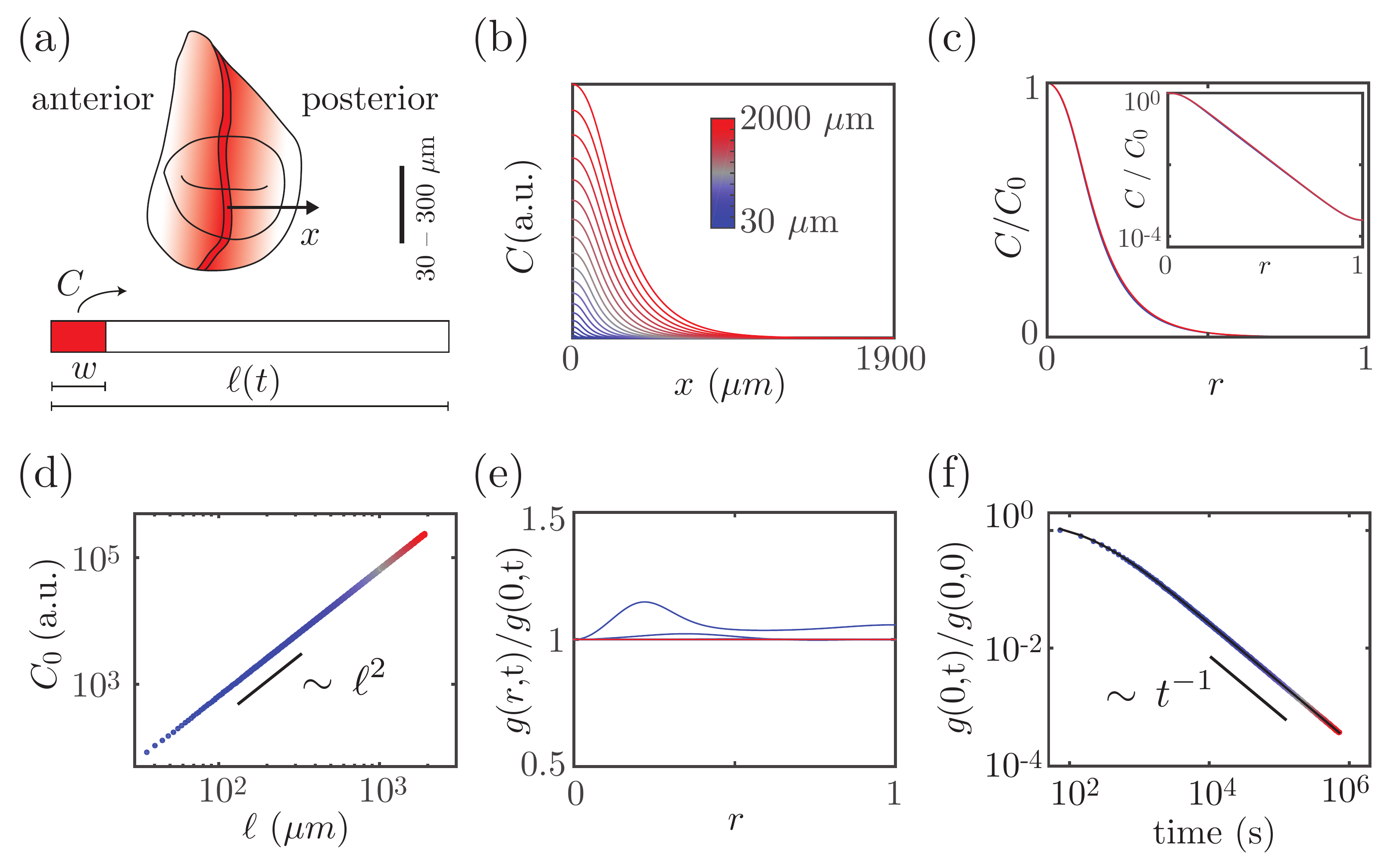}
\caption{Minimal model for growth control in biological tissues by scaling morphogen gradients.
(a) The wing imaginal disc of the fruit fly is a two dimensional epithelial sheet with a source releasing Dpp molecules (red) at the anterior-posterior (AP) compartment boundary (dark red). We consider a simplified morphogen system with a source at the left boundary. Panels (b)-(f) show numeric solutions of \cref{eq:cdiffgen,eq:growth} for $k{=}0$. Color code defined in (b) applies to all panels (in (c) and (e) most lines overlap). (b) Spatial profiles for morphogen concentration $C$ for different tissue lengths.
(c) Rescaled concentration profiles from (b) collapse on a master curve, thus showing scaling (inset: log-normal plot). (d) Amplitude $C_0$ of the concentration profile obeys a power-law relationship with tissue length $\ell$. (e) Self-consistently regulated growth becomes spatially homogeneous after an initial transient period. (f) Growth slows down inversly with time  (solid line: \cref{eq:g0t}). Parameters: $D{=}0.1\,\micron^2/\second$ \cite{Kicheva:2007bha}, $\nu{=}1\,\text{conc/s}$, $w{=}0.1\,\ell$, $\varepsilon{=}0.83$ \cite{Wartlick:2011by}, $\beta{=}2/(1+\varepsilon)$.}\label{fig:intro}
\end{figure}

In several model organisms it was observed that morphogen gradients scale proportionally with the size of the growing tissues, maintaining a constant shape \cite{Gregor:2005jn,Wartlick:2011by,BenZvi:2011jg,Bigfrogsmallfro:2009jy,BenZvi:2014ca,Werner:2015wm,Stuckemann:2017id}. 
Scaling of morphogen gradients and growth control has been studied in the fruit fly \textit{Drosophila melanogaster}, particularly in the precursor of the wing, the wing imaginal disc
\cite{Gregor:2005jn,Umulis:2010cd,Wartlick:2011by,BenZvi:2011jg}, see \cref{fig:intro}(a). 
Here, \textit{decapentaplegic} (Dpp) is one of the important morphogens implicated in tissue growth \cite{Capdevila:1994vt,Burke:1996tv,MartinCastellanos:2002vm,Schwank:2008fe,Schwank:2012gh,Wartlick:2012bn,Akiyama:2015hu,Harmansa:2015fs,Barrio:2017jt,Matsuda:2017fq,Bosch:2017ds}. 
Measurements at different stages of development revealed scaling of the Dpp concentration profile \cite{Wartlick:2011by,BenZvi:2011jg}, see numerical examples of pattern scaling in \cref{fig:intro}(b)-(c).
Several mechanisms have been proposed to explain scaling of the Dpp concentration profile with respect to compartment size \cite{BenZvi:2010es,Wartlick:2011by,Wartlick:2011fe,Averbukh:2014kb,Fried:2014fd,RomanovaMichaelides:2015hj}. One major class of mechanisms introduces an additional chemical species, termed expander, whose concentration depends on tissue size. It regulates morphogen dynamics and thereby scales its pattern \cite{BenZvi:2010es,Wartlick:2011by,Wartlick:2011fe,Averbukh:2014kb}.

Several mechanisms of growth control have been proposed \cite{Day:2000tu,Shraiman:2005hz,Rogulja:2005cy,Hufnagel:2007kb,AegerterWilmsen:2007cj,RomanovaMichaelides:2015hj}. One suggestion is that morphogen gradients control growth by a `temporal growth rule' \cite{Wartlick:2011by,Wartlick:2014cs}, 
where the local growth rate in the target tissue is set by relative temporal changes of the local morphogen concentration. 
This growth rule in conjunction with an expander mechanism for gradient scaling can account for the homogeneous growth observed in the wing imaginal disc \cite{Wartlick:2011by,Averbukh:2014kb} and may also apply to other tissues \cite{RomanovaMichaelides:2015hj,Fried:2016eh}. 
It has further been suggested that the temporal growth rule by itself could yield gradient scaling, 
without the need of an additional expander mechanism \cite{Averbukh:2014kb}.

In this letter, we propose a theoretical framework for the interplay between gradient scaling and growth control. 
In this framework, spatially homogeneous growth and exact scaling of chemical gradients both emerge as features of a critical point of the growth dynamics.
This approach provides a mechanism for the homogeneous growth and gradient scaling observed during the growth of the wing disc of the developing fly.

\paragraph{Morphogen dynamics and growth control.} 
We consider a minimal two-dimensional system with morphogen of concentration $C(\mathbf{x},t)$
as function of position $\mathbf{x}{=}(x,y)$ and time $t$. 
Morphogen dynamics is governed by local production in a specified source region $s(\mathbf{x},t){>}0$, by effective diffusion with diffusivity $D$, effective degradation with rate $k$, as well as by advection and dilution of molecules due to tissue growth.
Further, we consider a temporal growth rule by which the relative rate of change of the morphogen concentration controls the local rate $g$ of area growth \cite{Wartlick:2011by}, 
characterized by the dimensionless parameter $\beta$. 
Together, morphogen dynamics and growth control are described by 
\begin{align} \label{eq:cdiffgen}
D_t\,C=&\,\nabla\cdot\left(D\,\nabla\,C\right) -\left(k+g\right)\,C +s\,,\\
g=&\,\frac{1}{\beta}\,\frac{D_t\,C}{C}\,,\label{eq:growth}
\end{align}
\noindent where $\nabla$ is the gradient operator.
The convective time derivate $D_t{=}\partial_t+\mathbf{u}\cdot\nabla$ 
accounts for the local cell velocity field $\mathbf{u}(\mathbf{x},t)$ of the growing tissue,
which obeys $g{=}\nabla\cdot\mathbf{u}$ \cite{Wartlick:2011by}.

We consider a morphogen source aligned parallel to the $y$-axis with $s(x,t){=}\nu$ in the interval $0{\leq} x{\leq} w(t)$ 
and $s(x,t){=}0$ elsewhere, see dark red region in \cref{fig:intro}(a).
The width of the morphogen source is denoted $w(t)$ and $\nu$ is a production rate.
We consider morphogen profiles $C(x,t)$ and growth profiles $g(x,t)$ 
that only vary along the $x$-axis.
We choose reflecting boundary conditions at the domain boundaries, $x{=}0$ and $x{=}\ell$.
We account for a possible intrinsic anisotropy of tissue growth by the anisotropy parameter $\varepsilon{=}(\partial_y\,u_y)/(\partial_x\,u_x)$.
Thus tissue area scales as $A{\sim} \ell^{1+\varepsilon}$, where isotropic growth corresponds to $\varepsilon{=}1$.

\paragraph{Scaling of morphogen patterns.}
Scaling of concentration profiles is defined by the property that 
the time-dependent concentration $C(x,t)$ can be written as 
\begin{equation}
C(x,t)=C_0(t)\xi(x/\ell),
\label{eq:scaling}
\end{equation} 
where $\xi(r)$ with $r{=}x/\ell$ is a scaling function that characterizes a time-independent shape of the concentration profile and $C_0(t)$ is a time-dependent amplitude of the profile.
An example exhibiting this scaling property is shown in \Cref{fig:intro}(b)-(c).
It has been suggested that $C_0$ in \cref{eq:scaling} obeys a power law \cite{Wartlick:2011by} of the form
\begin{equation} 
C_0(t)\sim \ell(t)^q\,.
\label{eq:C0Lq}
\end{equation}
Scale invariance captured by scaling functions together with power laws often occur near critical points \cite{stanley1971}. This raises the question whether a critical point is underlying the scaling of morphogen pattens.

\paragraph{Growth control and conditions for scaling.}

Dynamic solutions of \cref{eq:cdiffgen,eq:growth} exist, which scale as described by \cref{eq:scaling,eq:C0Lq} and for which growth is homogeneous, as we show next.
This requires that the source width scales linearly with tissue length, $w(t){\sim} \ell(t)$.

Homogeneous growth with $g(x,t){=}\ghom(t)$ implies that the relative position $r{=}x/\ell$ of a material point does not change in time. 
In this case, the temporal growth rule \cref{eq:growth} simplifies to 
$\beta\, \ghom{=}\partial_t \ln(C_0)$. 
By definition, $\ghom$ is proportional to the relative change in tissue length, 
$\ghom{=}(1+\varepsilon)\, \partial_t \ln(\ell)$. 
Thus, we obtain the power law of \cref{eq:C0Lq} with exponent
\begin{equation} 
q=\beta\,(1+\varepsilon)\,.
\label{eq:scalingcond}
\end{equation}
This exponent takes a specific value, as we show now.
Combining \cref{eq:cdiffgen,eq:growth}, we have
\begin{equation}
\label{eq:comb13}
0=\,\nabla\cdot\left(D\,\nabla\,C\right) -\left[k+(1+\beta)\,g\right]\,C +s\,,
\end{equation}
which holds at all times.
For homogeneous growth, the time-dependent rate
\begin{align} 
k_g=k+(1+\beta)\,g \label{eq:kg} 
\end{align}
is position-independent, 
and the solution to \cref{eq:comb13} reads
\begin{align}
C(x,t)=\frac{\nu}{k_g}
\begin{cases}
1-\frac{\sinh{(\ell/\lambda-w/\lambda)}}{\sinh{(\ell/\lambda)}}\,\cosh{\left(\frac{x}{\lambda}\right)}&\text{, }x\leq w\\
\frac{\sinh{(w/\lambda)}}{\sinh{(\ell/\lambda)}}\,\cosh{\left(\frac{\ell-x}{\lambda}\right)}&\text{, } x>w\,,
\end{cases} 
\label{eq:Csolution}
\end{align}
where $\lambda{=}\sqrt{D/k_g}$ is a decay length.
The time-dependence of $C(x,t)$ arises from the time-dependencies of $\ell$, $w$, $\lambda$, and $k_g$. 
From \cref{eq:Csolution,eq:growth}, we find that growth is homogeneous if and only if concentration profiles scale.
This is the case if $\lambda{\sim}\ell$ and $w{\sim}\ell$.
Such scaling occurs if $k_g{\sim}\ell^{-2}$. 
Hence, $C_0{\sim} \nu/k_g$ obeys the power law \cref{eq:C0Lq} with $q{=}2$.
Together with \cref{eq:scalingcond}, we thus find that scaling can occur if the growth feedback parameter $\beta$
takes a critical value $\beta_c{=}2/(1+\varepsilon)$.

\paragraph{Growth dynamics and the effect of morphogen degradation.}

The time-dependence of homogeneous growth can be found 
using $k_g{\sim}\ell^{-2}$, \cref{eq:kg} and $\ghom{=}(1+\varepsilon)\,\dot{\ell}/\ell$, 
which together defines a differential equation for $\ell(t)$.
The solution depends on the value and time-dependence of the degradation rate $k$.
For the simple case $k{=}0$, a numerical solution to \cref{eq:cdiffgen,eq:growth} is shown in \cref{fig:intro},
highlighting that for $\beta{=}\beta_c$, after a short transient, growth is indeed homogeneous and concentration profiles scale.

We can obtain explicit expressions for the growth dynamics at this critical point $\beta{=}\beta_c$, 
revealing that growth is unbounded and the growth rate slows down as $t^{-1}$:
\begin{align}\label{eq:Ltkscale}
\ell(t)=&\,\ell(0)\,\big[1+2\,\ghom(0)\,t/(1+\varepsilon)\big]^{1/2}\,,\\
\ghom(t)=&\,\frac{\ghom(0)}{1+2\,\ghom(0)\, t/(1+\varepsilon)}\,,\label{eq:g0t}
\end{align}
see \cref{fig:intro}(f) and \cite{SM:2017}. 
Interestingly, the growth rate in the long-time limit $\ghom(t){\approx}(1+\varepsilon)/(2\, t)$ 
becomes independent of the initial conditions.

Exact scaling and spatially homogeneous growth is also found at $\beta{=}\beta_c$ for a finite but constant degradation rate $k{=}k_0{>}0$. 
In this case, the growth rate decays exponentially
\begin{align}
\ghom(t)=&\,\frac{\ghom(0)\,e^{-t/\tau}}{1+2\,\tau\,\ghom(0)(1-e^{-t/\tau})/(1+\varepsilon)}\,,\label{eq:g0texp}
\end{align}
with characteristic time scale $\tau{=}(1+\beta_c)(1+\varepsilon)/(2k_0)$. 
As a consequence, growth arrests at a final size $\ell^{\ast}$ \cite{SM:2017,Averbukh:2014kb},
\begin{equation} 
\ell^{\ast}=\ell(0)\,\big[1+g_0(0)(1+\beta_c)/k_0\big]^{1/2}\,.
\label{eq:ellstark0}
\end{equation}
Note that for $k_0{\rightarrow }0$, final size $\ell^{\ast}$ diverges as $\ell^{\ast}{\sim} k_0^{-1/2}$.

\begin{figure}[h!]
\includegraphics[width=0.9\linewidth]{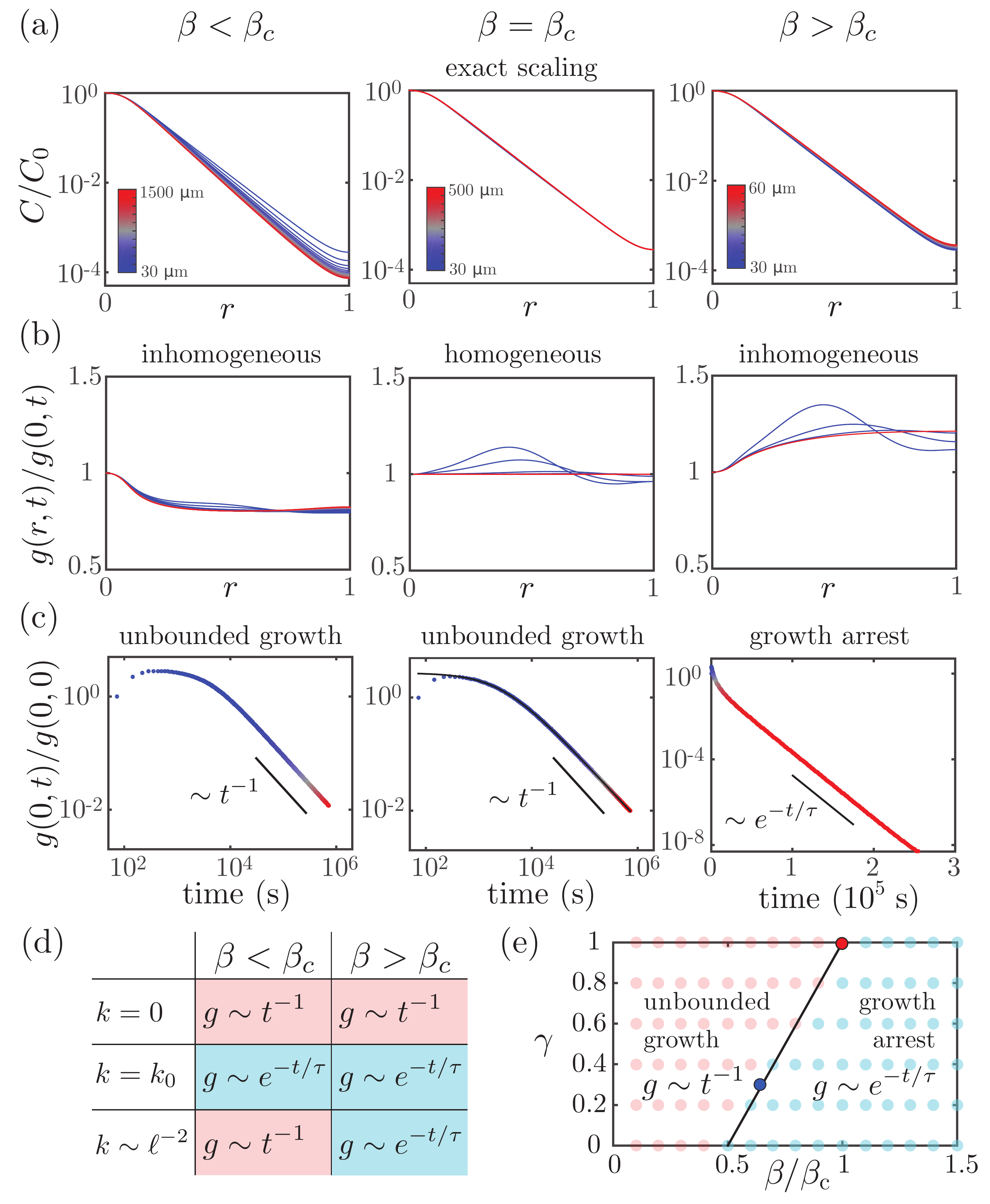}
\caption{Critical point and growth dynamics for $k{\sim} \ell^{-2}$. 
(a) Concentration profiles as a function of relative position $r{=}x/\ell$ for different tissue length (color code) and different values of $\beta$.
Scaling of the concentration profiles at the critical point with $\beta{=}\beta_c$ results in a collapse of the normalized concentration profiles for different tissue lengths. Above and below the critical point (here: $0.8\,\beta_c$, $1.2\,\beta_c$), deviations from scaling occur. (b) Growth rate profiles as a function of $r$ for different tissue length. At the critical point growth becomes homogeneous. (c) Growth rate as a function of time. For $\beta{>}\beta_c$, the growth rate decreases exponentially with time, while for $\beta{ \le} \beta_c$ a power-law behavior leads to unbounded growth (solid line: \cref{eq:g0t}). (d) Growth behaviors for super- and subcritical $\beta$ for different degradation scenarios. (e)
Different growth regimes as a function of the source scaling exponent $\gamma$ for $k{\sim}\ell^{-2}$. Regimes of unbounded growth (light red) and growth arrest (light blue) are separated by the line $\gamma{=}2\,\beta/\beta_c-1$ for $\gamma{<}1$. 
Numerical results (dots, see \cite{SM:2017}), critical point with $\gamma{=}1$ (red dot), parameters corresponding to fit
to experimental data shown in \cref{fig:exp} (blue dot). A constant source width corresponds to $\gamma{=}0$.
Parameters: $D{=}0.1\,\micron^2/\second$ \cite{Kicheva:2007bha}, $\nu{=}1\,\text{conc/s}$, $w{=}0.1\,\ell$  ($w{=}0.3 \,\micron\,(\ell/30\micron)^{\gamma}$ in panel (e)), $\varepsilon{=}0.83$ \cite{Wartlick:2011by}, $k\,\ell^{2}{=}9\,\micron^2/\second$. The color code defined in (a) also applies to (b) and (c).}\label{fig:theory}
\end{figure}

Next, we consider the degradation rate as a function of tissue length, $k{=}k(\ell)$, e.g. regulated by an expander \cite{BenZvi:2010es,Wartlick:2011by,Othmer:1980vg,Hunding:1988ve,Ishihara:2006to}. 
Let us consider the case of exact scaling of the degradation rate with tissue size in the form $k{\sim}\ell^{-2}$.
For $\beta{=}\beta_c$, we again find spatially homogeneous growth as well as exact pattern scaling, 
which is again described by \cref{eq:g0t,eq:Ltkscale}. In particular, growth is unbounded,
see \cref{fig:theory}. 
If, however, we add a small constant value $k_0$ to the degradation rate $k{-}k_0{\sim} \ell^{-2}$,
growth arrests at a finite size given by \cref{eq:ellstark0}.

These cases illustrate that at $\beta{=}\beta_c$, we can find either unbounded or bounded growth, depending on the behavior of the degradation rate $k$. In general, growth arrest can be observed if there exists a final size $\ell^{\ast}{>}\ell(0)$, for which $k_{g}(\ell^{\ast}){=} k(\ell^{\ast})$. 
This follows from \cref{eq:kg} \cite{SM:2017}.

\paragraph{A critical point of growth control.}
We now explore the behavior for $\beta{\neq}\beta_c$. 
In this case, the system does not exhibit exact pattern scaling and growth becomes spatially inhomogeneous, see \cref{fig:theory}(a)-(c) for an example. 
For $\beta{<}\beta_c$, $g(r,t)$ is decreasing with $r$, 
while for $\beta{>}\beta_c$, $g(r,t)$ is increasing with $r$, see \cref{fig:theory}(b) and \cite{SM:2017}.
As before, the growth dynamics depends on the degradation rate, see \cref{fig:theory}(d). 
Growth is always unbounded for $k{=}0$. 
For $k{=}k_0{>}0$, growth arrests at a finite final size $\ell^{\ast}$ for all values of $\beta$. 
In the case of $k{\sim} \ell^{-2}$, 
growth arrests for $\beta{>}\beta_c$ and the growth rate $g(t)$ decays exponentially with characteristic time $\tau$. 
The final size $\ell^{\ast}$ diverges as $\beta$ approaches the critical point $\beta_c$ from above. 
For $\beta{<}\beta_c$, growth is unbounded. 
Thus, $\beta{=}\beta_c$ exhibits distinct features of a critical point such as scale invariance of the concentration profile and divergent length scales. For $k{\sim}\ell^{-2}$ this critical behavior includes a transition between bounded and unbounded growth.

Only at the critical point, exact pattern scaling and homogeneous growth occurs.
However, in the vicinity of the critical point, 
patterns scale and growth is homogeneous to a good approximation, 
reflecting signatures of the critical point \cite{SM:2017}.
Interestingly, a control of the degradation rate by an expander molecule can maintain approximate scaling even away from the critical point if the growth rate is small compared to the degradation rate. In this case, $k_g{\approx} k{\sim} \ell^{-2}$, and growth inhomogeneities do not perturb scaling strongly, see \cref{fig:theory}(a).
Yet, even in this case of almost exact gradient
scaling, inhomogeneity of growth occurs depending on $\beta$, see \cref{fig:theory}(b).

\begin{figure}[t]
\includegraphics[width=0.94\linewidth]{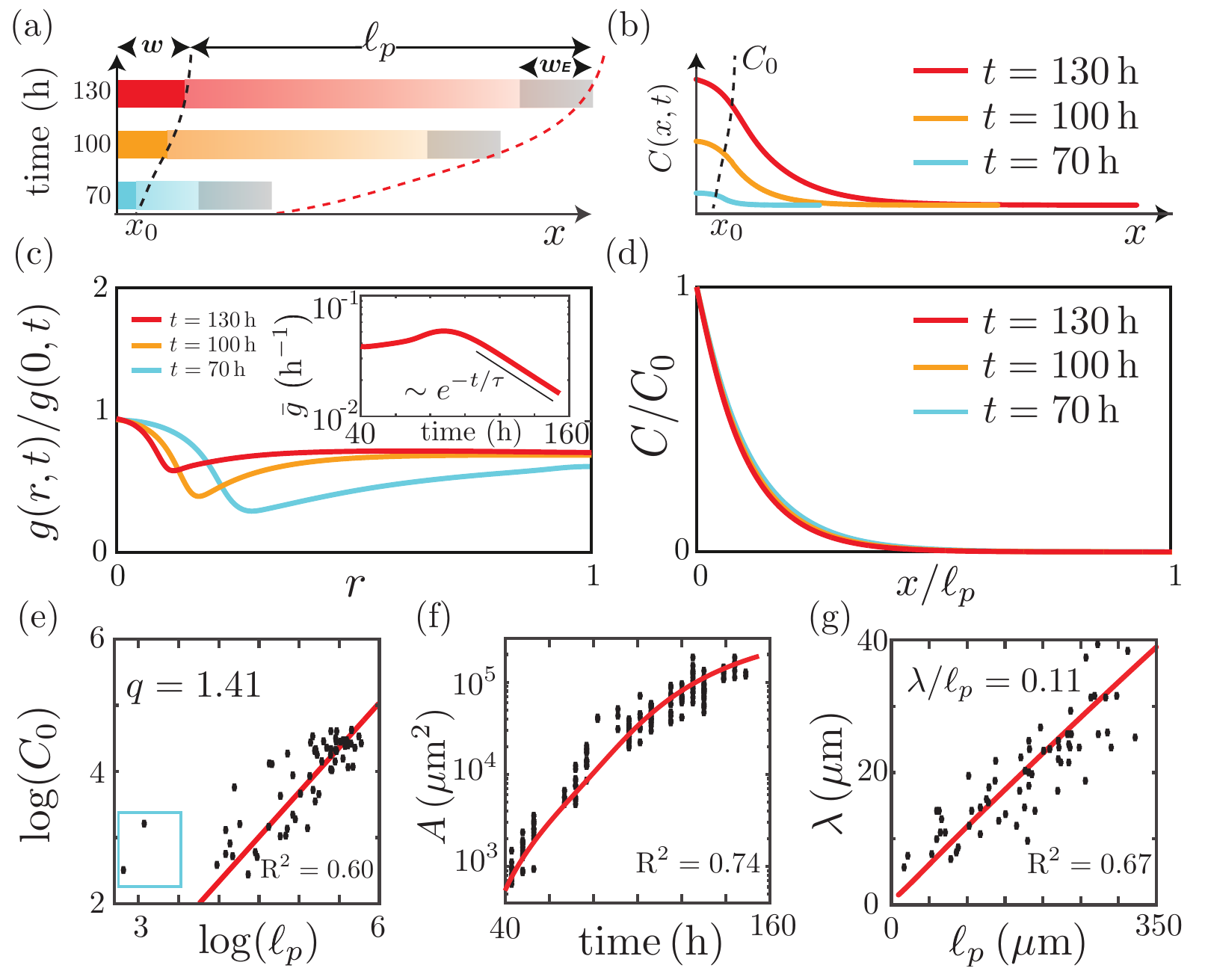}
\caption{Growth and gradient scaling in the fly wing.
(a) Schematic illustration of time-dependent morphogen profiles $C(x)$ in a growing posterior compartment of size 
$\ell_p$  regulated by an expander mechanism. The morphogen is produced in a source region of  
width $w$  that increases with tissue length $\ell$ as $w{=}w_0\,\ell^{\gamma}$. 
The expander is produced in a source of constant width $w_E$,  located at the posterior end, see \cite{SM:2017}. 
(b) Numerical solutions for morphogen profiles $C(x)$.
(c) Position and time dependence of local growth rates $g$. Inset: Average growth rate in the 
posterior compartment as a function of time. The growth rate relaxation time $48.2\,\text{h}$ is 
consistent with experiments \cite{Bittig:2009fs,Wartlick:2011by}. (d) Collapse of relative concentration profiles $C/C_0$ 
as function of relative position $x/\ell_p$ at different  times. (e)-(h) Comparison of experimental data (dots) 
\cite{Wartlick:2011by} and numerical solutions (solid lines).
(e)  Morphogen profile amplitude $C_0$ as a function of  posterior tissue size $\ell_p$.  
(f) Posterior tissue area $A$ as a function of time. (g) Decay length $\lambda$ of the morphogen profile in the posterior compartment as a function of $\ell_p$. Boxed data-points in (e) are excluded from the fits.  
Initial conditions: steady state of \cref{eq:cdiffgen}.  Parameters estimated
from experimental measurements: 
$D{=}0.1\,\micron^2\,\second^{-1}$ \cite{Kicheva:2007bha},
$\beta{=}0.7$,
$\varepsilon{=}0.83$ \cite{Wartlick:2011by}. Parameters estimated by a fit to the data: 
$\gamma{=}0.3$,
$w_0{=}5.75\,\micron^{1-\gamma}$,
$w_E{=}2.5\,\micron$,
$\nu/\nu_E{=}0.21$,
$k_E{=}5\cdot10^{-6}\,\second^{-1}$, 
$D_E{=}10\,\micron^2\,\second^{-1}$,
$\eta\,\nu_E^2{=}2.56\cdot10^{-11}\,\second^{-3}$.}\label{fig:exp}
\end{figure}

So far we focused on the case where the source width $w$ grows proportional to tissue length $\ell$. We now discuss situations where the source width is not proportional to tissue length. 
To simplify the discussion, we consider a source width $w{\sim}\ell^\gamma$ with $0{\leq} \gamma{<}1$, which interpolates between the cases of a constant source width ($\gamma{=}0$) and a source width proportional to tissue length ($\gamma{=}1$). Solving \cref{eq:cdiffgen,eq:growth} for different values of $\gamma{<}1$, we again find similar behaviors as described for $\gamma{=}1$. For example, two growth regimes can be distinguished,
depending on the value of $\beta$. For $\beta {<} (\gamma+1)/(1+\varepsilon)$, growth is unbounded and the growth rate as a function of time is well fit by a power law, while for $\beta {>} (\gamma+1)/(1+\varepsilon)$ growth is bounded and the growth rate is well fit by an exponential, see \cref{fig:theory}(e). Note that along the line $\gamma{=}2\,\beta/\beta_c-1$ we observe signatures of the critical point even for $\beta{<}\beta_c$, see \cref{fig:exp} and \cite{SM:2017}.

\paragraph{Homogeneous growth and gradient scaling in the wing imaginal disc of the fruit fly.}
Growth dynamics and spatial profiles of the morphogen Dpp have been quantified in the wing imaginal disc of the fruit fly \textit{Drosophila melanogaster}. Growth of the wing disc is approximately homogeneous and the growth rate decays exponentially with a time scale of 
$30{-}60\,\mathrm{h}$ \cite{Bittig:2009fs,Wartlick:2011by}.
Dpp profiles scale to a good approximation and their amplitude $C_0$ is well fit by a power-law relation with tissue area with exponent $\tilde{\beta}{=}q/(1+\varepsilon)$ ranging from $0.5$ to $0.7$ depending on the dataset \cite{Wartlick:2011by,SM:2017}.
Furthermore, homogeneous growth can be accounted for by the temporal growth rule \cref{eq:growth} with scaling Dpp 
profiles \cite{Wartlick:2011by}.
We show in \cref{fig:exp}(e)-(g) experimental data on tissue area $A$, tissue length $\ell$, decay length $\lambda$ and Dpp profile amplitude $C_0$ \cite{Wartlick:2011by} together with numerical values obtained by solving \cref{eq:cdiffgen,eq:growth}. 
This comparison shows that the continuum model can account for growth and Dpp concentration gradient dynamics in the wing imaginal disc.
The parameter values used in Fig. 3 are indicated in Fig. 2(e) as a blue dot.
Estimating the growth anisotropy $\varepsilon$ \cite{Bittig:2009fs,Wartlick:2011by} suggests that the growth parameter $\beta{\approx}0.7$ is smaller than $\beta_c{\approx}1.1$. Thus, the wing disc is not exactly critical. Deviations from criticality also arise because the source width in the wing imaginal disc 
increases less than linearly with tissue length. 
Experimental estimates locate $\gamma$ within the range $0.2{-}0.9$ \cite{SM:2017,Wartlick:2011by}, and our simulation fits experimental data of growth and morphogen dynamics with $\gamma{=}0.3$,
see \cref{fig:theory}(e) and \cref{fig:exp}(e)-(g). 
Therefore scaling and homogeneous growth are only approximate, and result as signatures of the nearby critical point.
Interestingly, the fly mutant Hh-CD2 differs from control animals in that its source width is constant \cite{Wartlick:2011by}. Hh-CD2 can be represented here by exponents $\gamma{=}0$ and $\beta{=}0.7$ \cite{SM:2017}, which locates its growth dynamics far from the boundary line between unbounded growth and growth arrest. 
From this observation we predict that scaling should be less precise and growth non-homogeneous for Hh-CD2 as compared
to control fly wings. Indeed, our analysis of Dpp-decay lengths is consistent with less precise scaling in Hh-CD2 \cite{SM:2017}.

\paragraph{Conclusion.} 

We presented a theory for self-organized growth of tissues regulated by a dynamic morphogen profile and a temporal growth rule.
We find that both exact scaling of the morphogen profile and homogeneous growth are mutually dependent and arise as features of a critical point.
We determine a concise condition for scaling and homogeneous growth in terms of a critical feedback strength. We reveal characteristic features of the presented mechanism: First, the amplitude of morphogen profiles obeys a power-law relationship with tissue length. Second, there exist distinct regimes of growth arrest and unbounded growth in which spatial profiles of growth differ. Third, scaling itself is independent of many details of the dynamic equations if the system is close to criticality.
In particular, scaling does in principle not require an expander mechanism and could occur even in the 
absence of a feedback on tissue length \cite{Averbukh:2014kb}. However, an expander can alter the growth dynamics.
Note that an expander regulation that provides the relation $k{\sim}\ell^{-2}$ leads to unbounded growth at the critical point. Reliable growth termination can be achieved by an offset in the scaling relation, e.g. $k{-}k_0{\sim}\ell^{-2}$. Such behavior could occur for example in the case of delayed expander regulation.

We applied our theory to the dynamics of morphogen gradients and growth during the development of the wing imaginal discs of the fruit fly. 
Chosen parameters, which are consistent with previous experiments, correspond to $\beta{<}\beta_c$, but are close to the boundary in parameter space separating bounded from unbounded growth (\cref{fig:theory}(e)). We find that nonlinear scaling behavior of the Dpp source, as quantified in \cite{Wartlick:2011by}, may place the wing disc in the regime of bounded growth even for a super-critical growth parameter.
Our work suggests that in the wing imaginal disc an expander 
mechanism ensures that growth arrests, while the scaling of Dpp profiles and the spatial homogeneity of growth
result as robust signatures of a critical point.
The framework presented here could be applied to other systems, such as the eye imaginal
disc of the fly, which is an example of a non-stationary Dpp-source that orchestrates growth \cite{Wartlick:2014cs}.

\paragraph{Acknowledgments.} 
We thank Maria Romanova-Michaelides and Zena Hadjivasiliou for discussions.
D.A.H., F.J. and M.G.G. acknowledge support from the DIP of the Canton of Geneva, SNSF, the SystemsX epiPhysX grant, the ERC (Sara and Morphogen), the NCCR Chemical Biology program and the Polish-Swiss research program.
S.W. and B.M.F. acknowledge support from the German Federal Ministry of Education and Research (BMBF), Grant No. 031 A 099, and DFG through the Excellence Initiative by the German Federal and State Governments (cluster of excellence cfaed).

D.A.H. and S.W. contributed equally to this work.

\bibliographystyle{apsrev4-1}

%\bibliography{scaling}

%

\end{document}